\documentstyle[12pt,preprint,prc,aps]{revtex}

\begin{document}

\draft

\title{
What Can Be Learned by   Measuring the Fluxes of the $^7{\rm Be}$ and the
$pep$ Solar Neutrino Lines ?}

\author{J.N. Bahcall and P.I. Krastev}
\address{School of Natural Sciences, Institute for Advanced 
Study\\
Princeton, NJ 08540\\}
\maketitle
%\newpage
\begin{abstract}
{Measurements of the interaction rates of 
the solar neutrino lines of $^7$Be  and $pep$ can be used,
independent of solar models, to test whether electron flavor is
conserved, to determine survival probabilities of electron-type neutrinos
at specific 
energies, and to test for the existence of sterile neutrinos. 
We present analytic descriptions of these tests.  We also illustrate
by numerical simulations, 
assuming matter-enhanced and vacuum neutrino oscillations, 
what measurements of  solar neutrino lines can teach us 
about neutrino  masses and mixing angles.
}

\end{abstract}

\section{Introduction}
\label{introduction}

Figure~\ref{recoilspectrum} shows the electron recoil spectrum that we
calculate  for neutrino-electron
scattering.
Although, as Figure~\ref{recoilspectrum} makes clear, 
the sun is predicted to 
produce important sources of low-energy 
neutrinos with energies of order an MeV or  less, there are no
operating detectors that can 
measure individual neutrino energies in this low energy range.
In particular, there is currently no way to isolate experimentally the fluxes of the
predicted strong solar neutrino lines.

We urge readers who are familiar with solar neutrino research
to turn immediately to Section~\ref{summary}, which contains a concise
summary and discussion of our  analysis of what can be learned about
neutrino properties from 
studying solar neutrino lines. We do not discuss here what can be
learned about the solar interior from studying neutrino lines\footnote{In
ref\cite{bahcall94}, it is shown that a measurement of the energy
shift of the $^7$Be solar neutrino 
line is equivalent to a measuremnt of the central
temperature of the sun and a measurement of the energy profile of the
$^7$Be line will determine the temperature profile of the solar
interior.}. 

Continuum neutrino sources, principally
neutrinos from $^8$B decay and from the $pp$ reaction, are believed to
be the major contributors to the 
four pioneering solar neutrino experiments: chlorine\cite{chlorine},
Kamiokande\cite{kamiokande}, GALLEX\cite{gallex}, and SAGE\cite{sage}.
Moreover,  the two next-generation  experiments,
Superkamiokande\cite{superkamiokande} and SNO\cite{sno}, are both   
sensitive only to the  neutrino continuum  from 
$^8$B decay.

The four pioneering 
detectors have established experimentally that the sun shines
by nuclear fusion reactions among light elements. 
Table~\ref{snudata}
summarizes the results of these  experiments.

Because 
the observed rates are lower than the predicted rates, the results
from the operating  experiments have led to
a number of suggestions for  new
particle physics.  
In these particle physics scenarios, something causes a fraction of electron
type neutrinos to disappear, or change their flavor,  
after they are created in the center of the sun.
All of the particle physics solutions of the solar neutrino problem
predict a survival probability, the probability that an electron-type
neutrino remains an electron-type neutrino, that is different from
unity.  In contrast, the 
survival probability is  equal to unity  
in  the simplest version of standard electroweak 
theory\cite{GWS} .

Two new experiments, SNO and Superkamiokande, 
were designed with the goal in mind of  establishing definitively 
 if physics beyond the standard electroweak model 
is required to explain the results of solar neutrino experiments.
Moreover, these 
new experiments will have the potential to determine the total
flux (independent of flavor) of $^8$B solar neutrinos, thereby testing
the prediction by solar models of the flux for this rare mode of
neutrino production.

The most plausible particle physics explanations, resonant matter
oscillations\cite{msw,hata93,bk96} (the
Mikheyev-Smirnov-Wolfenstein, or MSW, effect) and vacuum neutrino
oscillations\cite{vac}, both predict a strong energy
dependence for the survival probability. The form of the energy  dependence
 is determined by the specific parameters used in the adopted
oscillation scenario.  Other suggested new particle physics
explanations that predict a strong energy dependence for the survival
probability include
neutrino decay \cite{bah72,ber87}, non-standard electromagnetic 
properties \cite{vol86,lim88,akh88}, neutrino violation of the 
equivalence 
principle \cite{gasp88}, and 
supersymmetric flavor-changing neutral currents
\cite{rou91,guz91}. Many of the relevant papers (and further references)
are reprinted in \cite{30yr}.

We explore in this paper what can be learned about neutrino physics
by performing experiments with solar neutrino lines. 

Measurements with solar neutrino lines have the advantage that the
predictions of 
particle physics models are more specific for a line source
than they are for a continuum source.
Measurements of continuum interaction 
rates determine a weighted average of what
happens to neutrinos of different energies.
Moreover, 
if a neutrino line is detected in two ways (e.g., by
neutrino-electron scattering and by neutrino absorption)
then the survival
probability at a specific energy can be determined empirically,
independent of any solar physics.  At any energy, 
a  measured value for the survival
probability that is significantly different from unity would
constitute evidence for electron flavor non-conservation.

There are two nuclear reactions that are predicted to emit 
detectable
numbers of solar neutrinos
with specific energies, {i.e.,} 
neutrino lines.\footnote{Although the fluxes from seven solar neutrino 
lines have been
estimated\cite{bahcall90}, only the fluxes from the $^7$Be and the $pep$
neutrino lines are expected to be large enough to be measurable in solar
neutrino experiments that are currently feasible.}  The more frequent of these
reactions produces $^7$Be neutrinos via

\begin{equation}
{\rm ^7Be} ~+~ e^- ~ \rightarrow ~{\rm ^7Li} ~+~ \nu_e.
 \label{be7reaction}
 \end{equation}
Reaction~(\ref{be7reaction}) produces, according to the standard solar
model\cite{bahcall89,BP95}, 
a total neutrino  flux at earth of $5\times10^9~{\rm cm^{-2}s^{-1}}$, 
 89.7\% of the neutrinos having, in the laboratory, 
an energy of $E_\nu = 0.862$ MeV and
the other 10.3\% have $E_\nu = 0.384$ MeV .
The branching ratio of 9:1 is determined by nuclear physics and is the
same in the laboratory and in the solar interior.
The two $^7$Be  lines can, in principle, be used to perform a unique test of the
existence of sterile neutrinos \hbox{(see Section~\ref{teststerile})}.

The $pep$ neutrinos are created by the reaction

\begin{equation}
p ~+~ e^- ~+~p \rightarrow ~^2H ~+~ \nu_e.
 \label{pepreaction}
 \end{equation}
Reaction~(\ref{pepreaction})
produces neutrinos of energy $E_\nu = 1.442$ MeV, with 
a standard solar model flux of $1.4\times10^8~{\rm cm^{-2}s^{-1}}$.
In the standard solar model calculations,
the total flux of $^7$Be neutrinos is about $35$ times larger than the
total flux of $pep$ neutrinos.  If neutrino oscillations occur, the 
predicted  standard model ratio of $^7$Be to $pep$ neutrino fluxes may be
much reduced.

 So far, 
BOREXINO\cite{borexino}, which will observe neutrino-electron
scattering,
is the only detector  in an advanced stage of development
that is being constructed with 
a goal  of isolating events from a solar neutrino 
line.  Two other experiments are being developed, HELLAZ\cite{HELLAZ}
and HERON\cite{HERON}, which have the potential to detect solar
neutrino lines via neutrino-electron scattering.
Most recently, 
a Ga-As detector of low energy neutrinos has been
proposed\cite{gaas}. 
This detector could potentially measure 
the $pep$ and
the $^7$Be neutrino lines  by neutrino-electron scattering and,
very importantly, also by neutrino absorption (which would determine the
charged current rate).
For all of these experiments, good energy resolution will be required
in order to separate the solar neutrino lines from continuum solar
neutrino sources (cf. Figure~\ref{recoilspectrum}) 
and from background events.

We  begin, in Section~\ref{nusc}, by
discussing neutrino-electron scattering experiments.
We calculate
the predicted electron recoil spectrum in
neutrino-electron scattering experiments for four different neutrino
oscillation scenarios and for the standard model (no oscillations,
standard solar model). We  then show to what extent 
measurements of the scattering rate for neutrino lines can be used to
help determine neutrino masses and mixing angles.  We calculate 
how much additional information can be gained by measuring
both the $^7$Be and the $pep$ neutrino lines, rather than 
concentrating (as originally planned in the BOREXINO experiment) on
the $^7$Be line. 

Next we  demonstrate in Section~\ref{absorptionplus} 
how  survival probabilities at a specific energy 
can be measured if a neutrino line
is studied both by a neutrino absorption experiment and by a
neutrino-electron scattering experiment. 
We present a simple formula, Eq.~(\ref{survival}), for the survival
probability at the energy of the neutrino line; this formula is
independent of all solar physics.
We show in Section~\ref{ncexperiments} 
how neutral current experiments, when
combined with either absorption or neutrino-electron scattering
experiments can be used to determine the survival probability.
 In 
Section~\ref{modelindependent}, we focus on 
 the model-independent inferences
that are possible if both neutrino absorption and neutrino-electron
scattering are measured.  
We show that, in a two-flavor scenario, the neutrino mixing angle and
mass difference can be determined with reasonable accuracy if the 
absorption and neutrino-electron scattering rates are measured for
both the $^7$Be and the $pep$ lines. In Section~\ref{sterile}, we
show how studies of solar neutrino lines can  help answer
the question: Do sterile neutrinos exist?  
We summarize and discuss in Section~\ref{summary} 
the results obtained in this paper.

For the interested reader, we note that Bilenky and
Giunti\cite{bilenky93} have discussed, in a series of original and
stimulating papers, the possibilities for using experiments
that study the ${\rm ^8B}$ continuum solar neutrinos
to determine survival probabilities and to test for the existence of
sterile neutrinos.

How can we assess what
will be learned from different experiments without knowing 
which solution of the
solar neutrino problem Nature has chosen? 
We must adopt some tentative model for how neutrinos behave in order to
proceed. 
We assume successively the validity of either 
the small mixing angle (SMA)  or the large mixing angle
(LMA) 
MSW  solutions\cite{msw}, or the vacuum (VAC) neutrino oscillation 
solution\cite{vac}. 
We also assume the
correctness of the four operating solar neutrino experiments, which
fix the best-fit neutrino mixing parameters, $\Delta m^2$ and $\sin^2
2\theta$. Using these best-fit parameters, we compute the expected
event rates in  future experiments.  Assigning random errors of
plausible size to future measurements we analyze together the four
pioneering experiments that have been performed and the simulated new
experiments. 
We establish 95\% confidence limits on neutrino 
parameters that are consistent
with the four operating experiments, and with simulated results of
future experiments, using the techniques described previously 
\hbox{in\cite{bk96}}.

\bigskip\bigskip

\section{Neutrino-Electron Scattering Experiments}
\label{nusc}

 In this section, we determine by how  much measurements of the rates of 
neutrino-electron
scattering by $^7$Be or
 $pep$ neutrino lines can 
reduce the allowed regions
in the neutrino mass versus neutrino mixing angle plane. 
The neutrino-electron scattering  reaction can be represented by the 
equation 

\begin{equation}
\nu ~+~ e \rightarrow ~\nu^\prime ~+~ e^\prime.
 \label{reaction}
 \end{equation}

Neutrino-electron scattering experiments are sensitive to both 
charged current
(i.e., only $\nu_e$) and neutral current (i.e., all neutrino flavors)
interactions.  For the $^7$Be line, the ratio of the total electron
neutrino scattering cross section  
to the total neutral current  cross section is 4.53.  For
the $pep$ line, the corresponding ratio of the cross sections is 
4.93.  
All of the results given in this paper include radiative corrections
according to the perscription of reference\cite{bks95}.

Table~\ref{snuscenarios} gives the recoil electron event rates in SNU
predicted by different  solutions of the solar neutrino problem for 
individual neutrino sources.  The neutrino oscillation parameters used
in Table~\ref{snuscenarios} were found by requiring that $\chi^2$ be a
minimum for the four experimental results described in
Table~\ref{snudata}.  The neutrino parameters differ slightly from
those found in reference \cite{bk96} because in the present paper we
include radiative corrections for neutrino-electron scattering.  The
standard model fluxes are from reference \cite{BP95}.

\subsection{Recoil Electron Spectrum}
\label{nuscrecoil}

Figure~\ref{recoilspectrum} shows the calculated energy spectrum of 
the recoil electrons for five
 conceivable scenarios: the standard solar model and no neutrino
oscillations (indicated by the solid line), the SMA MSW solution 
(the dotted
line), the LMA MSW solution (the short dashed line), and the 
vacuum neutrino oscillation solution (long dashed line).  We also show
the result for the 
extreme case\cite{bethe39,bfk96} in which the solar
luminosity is assumed to be  produced by CNO reactions 
(the dot-dash curve).

The vertical axis of Figure~\ref{recoilspectrum} gives the calculated
event rate in SNU per MeV, where \hbox{$1$ SNU} is $10^{-36}$ 
interaction per
target electron per sec.
The horizontal axis, ${\rm T_{\rm e}}$, is the kinetic energy of
the recoil electron.  
Radiative corrections\cite{bks95} are included in the cross section calculations.

For neutrino-electron scattering,
$1$ SNU is approximately $2.6$ events per $100$ tons per day (for a
target material in which the mass number, $A$, equals twice the atomic
number, $Z$).  

If neutrino oscillations do not occur, then the computed 
shape of the recoil electron spectrum for standard solar model fluxes 
has prominent sharp shoulders at the
maximum kinetic energies of the $^7$Be and the $pep$ neutrino 
lines, respectively, i.e., at 
$1.225$ MeV and $0.665$ MeV. These features are apparent for the
standard model spectrum (solid line) 
in Figure~\ref{recoilspectrum}. The large continuum 
contribution from $pp$ neutrinos
is confined to energies less than $0.261$ MeV.  The $^{13}$N and
$^{15}$O continuum neutrinos can produce maximum electron recoil kinetic
energies of $0.988$ MeV and $1.509$ MeV, respectively. The rare ${\rm
^8B}$ neutrinos produce a low-level continuum up to 14~MeV.

As shown by many authors \cite{gelb92}, the neutrino-electron scattering rate 
may be much lower than in the standard solar model
predictions if neutrino oscillations occur.

For the CNO solution, the predicted event rates for energies less than
$1.5$ MeV are larger than if
standard model neutrino fluxes are assumed\cite{bfk96}.  In the energy
region in which the $^7$Be line produces electron recoils, the
calculated event rate is larger by typically a factor of about 2.5  
than what is expected from the
standard solar model.  Even greater enhancements in the predicted
event rates, a factor of 7 or
more, are implied by the  CNO scenario in the region $0.7$ MeV 
to $1.2$ MeV,
 in which the electron recoils from scattering by the $pep$ neutrino 
line are found.

\subsection{Allowed Regions of Neutrino Parameter Space}
\label{nuscallowed}

We begin this subsection  by stating an important result  
that refers to all three of the neutrino
oscillation scenarios.  By a series of detailed calculations, 
we have found  that 
a  measurement of the electron scattering rate of either the $^7$Be 
$0.862$ MeV line or the
$pep$ line to an accuracy of
$10$\% would, in conjunction with the four operating experiments, 
essentially eliminate the two competing oscillation 
scenarios that
are assumed, for purpose of that particular  simulation, not to be 
correct. 
Thus, for example, 
if the SMA scenario is assumed to be 
correct and either the $^7$Be or the $pep$ line is
measured to a $10$\% accuracy, 
then both the LMA and the vacuum oscillation
scenarios will be ruled out.  
Experiments with a much improved accuracy of $5$\% do not
provide significantly more stringent constraints on allowable
oscillations hypotheses.

Neutrino-electron scattering experiments
can determine which, if any, neutrino 
oscillation scenario is correct.

Table~\ref{bestfit} gives the predicted results for future experiments
on solar neutrino lines that are implied by the best-fit oscillation
descriptions of the four pioneering solar neutrino experiments.  The
predicted event rates and confidence limits for neutrino-electron
scattering were computed using the techniques of reference\cite{bk96} and
the standard model fluxes of reference\cite{BP95}.

The results of our more specific calculations for neutrino-electron 
scattering experiments are summarized in Figure~\ref{escallowed}a-c.
  The regions in the mass versus mixing angle plane that are  
allowed by the four operating solar neutrino experiments are 
   delineated by the solid lines in Figure~\ref{escallowed},
 which are taken from \cite{bk96}. 
Assuming $10$\% experimental errors for future measurements, 
   the dotted lines show the smaller allowed regions if a 
   measurement of the $^7$Be
   line rate is made.  If a $pep$ measurement is also made, the 
allowed regions
   are reduced still further to the regions indicated by the dashed lines.
In all cases, we determine the 95\% C.L. by requiring that the
boundaries of the allowed region satisfy
$\chi^2 = 5.99 + \chi^2_{\rm min}$. 

 The top panel, Figure~\ref{escallowed}a, was constructed assuming 
the correctness of the
best-fit small mixing angle solution of the solar neutrino problem. 
The dark circle shows
the position in the mass and mixing angle plane of the best-fit
solution and the
dark line shows the 95\% confidence limits of the parameters
determined by a 
$\chi^2$ fit to the results of the four pioneering solar neutrino 
 experiments.  
With the four  published experimental results, 
the large mixing angle solution is 
also allowed.

A measurement of the scattering rate of neutrinos from the 
$0.862$ MeV $^7$Be line
would,
with the given assumptions, eliminate the large mixing angle solution and
reduce significantly the allowed area of the small mixing angle solution.
The additional  measurement of the $pep$ line would reduce only 
slightly the
allowed region.

The
middle panel of Figure~\ref{escallowed}b refers to the case in which
the 
large mixing angle
solution is  correct. 
The allowed region for the SMA solution with just the four
operating experiments is slightly larger in Figure~\ref{escallowed}b
than in Figure~\ref{escallowed}a, because for the LMA solution adopted
in the middle panel $\chi^2_{\rm min, ~LMA} = 2.5$ whereas for the SMA
solution the fit is much better, $\chi^2_{\rm min, ~SMA} = 0.3$. 

A $^7$Be
measurement would reduce significantly the allowed range of LMA 
parameters and
almost entirely eliminate the permitted SMA parameter space,
as can be seen from Figure~\ref{escallowed}b.
The vacuum oscillation solution would also be ruled out.
Adding a
measurement of the $pep$ line would, in this case, significantly 
reduce the
remaining parameter space.  All that would be left would be a 
relatively small
region  surrounding  the best-fit LMA solution.

Finally, we show in Figure~\ref{escallowed}c the potential results 
assuming the correctness of
the vacuum neutrino oscillations.  
The best-fit  value of $\chi^2_{\rm min, ~VAC} = 2.5$.
In this case, the $^7$Be 
line measurement
greatly reduces the allowed parameter space for the vacuum 
oscillations and
completely eliminates the SMA and LMA MSW solutions.  The $pep$ 
measurement
makes a further dramatic reduction of the allowed parameter space, 
centering
the overall allowed region on a small area closely surrounding the 
best-fit
point determined from the four existing experiments.

\subsection{Summary of Potential of Neutrino-Electron Scattering
Experiments}
\label{nuscpotential}

We conclude this section with a brief summary of what can be learned from
neutrino-electron scattering experiments using the $^7$Be and 
$pep$ neutrino
lines.  
The electron recoil spectra expected, see Figure~\ref{recoilspectrum},
are different depending upon whether the sun shines by $pp$ or
CNO fusion reactions.  If the CNO cycle is the dominant source of
energy generation, the expected event rate is larger in the region in
which the electron recoil energy  is less than $1.5$ MeV and the shape
and energy span 
of the recoil electron energy spectrum is different,  than would be
expected if $pp$
reactions are most important source of solar energy generation.  
 
An accurate  measurement of the scattering rate of the 
$^7$Be or the $pep$ line would allow only one of the three popular
neutrino oscillation scenarios.  If the $^7$Be line is measured, then the
additional measurement of the $pep$ line would provide a major further
reduction in the allowed range of neutrino parameters if either 
the LMA or the
vacuum oscillation solution is correct.  If the SMA solution has 
been chosen by
Nature, then the $pep$ line may not add much additional information.

\bigskip\bigskip
\section{Absorption Plus Electron Scattering Experiments}
\label{absorptionplus}

In this section, we show how a  neutrino absorption
(charged-current) experiment, when combined with an electron scattering
experiment, makes possible the measurement of the  neutrino
survival probability at a specific energy.
Relevant neutrino-electron scattering 
experiments include BOREXINO\cite{borexino}, 
HERON\cite{HERON},  and HELLAZ\cite{HELLAZ}, while 
Ga-As\cite{gaas}, and  $^7$Li\cite{lithium} are 
candidates for an absorption
detector of the $^7$Be and $pep$ neutrino lines.
One advantage of a lithium detector in this connection is that the
absorption cross sections are large and are accurately
known\cite{bahcall78}  because
the inverse reaction ($^7$Be electron capture) is well studied in the
laboratory. 

In \ref{specific}, we present the formulae that determine the
survival probability in terms of the measured rates of the
absorption and scattering experiments.  In the following subsection,
\ref{bothallowed}, we present a graphical description of the allowed
regions in the absorption-scattering plane that are permitted by the
four operating solar neutrino experiments. Performing both an
absorption and a scattering experiment using a neutrino line 
selects a unique point in the absorption-scattering plane (or,
with experimental errors, a unique region) that determines the survival
probability at the energy of the line.

In the formulae presented in this section and in
Sections~\ref{ncexperiments} and \ref{modelindependent},
we assume that there are no sterile neutrinos.  In Section~\ref{summary},
we generalize the results to the case in which sterile neutrinos
exist.

\subsection{The Measurement of  Survival Probabilities at a Specific
Energy }
\label{specific}

Consider  an electron-type neutrino with energy $E_\nu$
that is created in the interior of the sun.  We denote by $P$ the 
probability that the neutrino remains an electron-type neutrino when
it reaches a detector on earth, i.e., 
$P = P(\nu_e \rightarrow \nu_e; E_\nu)$.  
In the literature, $P$ is usually referred to as a ``$\nu_e$ survival
probability.'' 
The rate per target atom for the charged-current
(absorption) reaction at energy $E_\nu$ may be written

\begin{equation}
R_{\rm abs} = \sigma_{\rm abs} P \phi,
\label{CCrate}
\end{equation}
where $\sigma_{\rm abs}$ is the absorption cross section, 
and $\phi$ is the total flux  of neutrinos of energy $E$
created in the sun.  In what follows, we suppose that $P$
is averaged over the neutrino 
production region in the interior of the sun.
The rate per target electron for the electron scattering reaction is

\begin{equation}
R_{\rm esc} = \left[\sigma_{\rm esc}(\nu_e) - \sigma_{\rm
esc}(\nu_{\rm x}) \right] P \phi ~+~
\sigma_{\rm esc}(\nu_{\rm x}) \phi,
\label{escrate}
\end{equation}
where $\sigma_{\rm esc}$ is the electron scattering cross section
and $\nu_{\rm x}$ is any normalized linear superposition of $\nu_\mu$ and
$\nu_{\tau}$. 

Combining Eqs.~\ref{CCrate} and \ref{escrate}, we obtain an explicit
expression for the survival probability for electron-type neutrinos of
energy $E$:

\begin{equation}
P = \frac{\sigma_{\rm esc}\left(\nu_{\rm x}\right)R_{\rm abs}}
{\sigma_{\rm abs}R_{\rm esc} -
\left[\sigma_{\rm esc}\left(\nu_e\right) - 
\sigma_{\rm esc}\left(\nu_{\rm x}\right)\right]R_{\rm abs} }
\label{survival}
\end{equation}
Equation~(\ref{survival}) could  be used to determine, independent of any
solar physics,  the survival
probability at a specific energy for  neutrinos produced in
either the $^7$Be or the $pep$ line.

How well do experiments with specific uncertainties determine the
survival probability?  This question is answered by 
Eq.~(\ref{partialscabs}), which is  shown
below.

\begin{equation}
- \frac{\partial ln P}{\partial ln R_{\rm esc}} = + \frac{\partial ln
P}{\partial ln R_{\rm abs}} = \frac{\left[ \left(
\sigma_{\rm esc}(\nu_e)  -\sigma_{\rm esc}(\nu_{\rm x}) \right)P ~+~
\sigma_{\rm esc}(\nu_{\rm x}) \right]}{\sigma_{\rm esc}(\nu_{\rm x})} .
\label{partialscabs}
\end{equation}

To estimate the accuracy with which
$P$ is determined by a given pair of experiments, one inserts the
best-estimate of $P$ obtained from Eq.~(\ref{survival}) in the 
right hand side of Eq.~(\ref{partialscabs}).
The fractional uncertainty in the inferred survival probability for
given experimental errors can then be determined by multiplying
Eq.~(\ref{partialscabs}) by the fractional uncertainty, $\Delta R_{\rm
esc}/R_{\rm esc}$,  in the measured neutrino-electron scattering rate
or by the fractional uncertainty, $\Delta R_{\rm
abs}/R_{\rm abs}$, in the measured neutrino absorption rate.  

The
uncertainty in the experimentally determined survival probability 
depends only upon the survival probability itself and upon the ratio
of neutrino-electron scattering cross sections, 
$ \sigma_{\rm esc}\left(\nu_e\right)/\sigma_{\rm
esc}\left(\nu_{\rm x}\right)$. For a very small inferred survival
probability, the fractional uncertainty in the probability that
results from a measurement with a specified fractional error is equal
to that fractional error.  For survival probabilities close to unity,
the fractional error in the inferred survival probability is amplified
by a factor of $ \sigma_{\rm esc}\left(\nu_e\right)/\sigma_{\rm
esc}\left(\nu_{\rm x}\right)$ relative to the error in the measurement.

\subsection{Allowed Parameter Regions if Electron Scattering and
Absorption are Measured}
\label{bothallowed}

We can rewrite Eq.~(\ref{survival}) as a linear relation between the
neutrino-electron scattering rate and the charged current rate.
Dividing Eq.~(\ref{CCrate}) and Eq.~(\ref{escrate}) by the standard
model expectations, one finds

\begin{equation}
\frac{R_{\rm esc}}{R_{\rm esc, SSM}} = \frac{R_{\rm abs}}{R_{\rm abs,
SSM}} + \frac{\sigma_{\rm esc}(\nu_{\rm x})(1 - P)}{\sigma_{\rm esc}
(\nu_e)} .
\label{linear}
\end{equation}

Figure~\ref{belinear} displays in the electron-scattering versus charged
current plane the linear relation between the two measurable event
rates.
The upper panel of Fig.~\ref{belinear} shows the 95\% C.L. regions
that are allowed by the four operating experiments for the SMA and the
LMA MSW solutions.  The two sets of MSW solutions overlap slightly but
are well separated from the predictions of the standard model.  The
lower panel of Fig.~\ref{belinear} shows the relatively larger range 
 that is allowed  by the vacuum oscillation solutions.

Figure~\ref{peplinear} displays similar information for the $pep$
line.
Note that the upper panel of Fig.~\ref{peplinear} shows that the SMA and
the LMA solutions are distinguishable if both the charged current and
the electron scattering rates are measured for the $pep$ line. The
allowed range of vacuum solutions is, however, very large, as is shown
by the lower panel of Fig.~\ref{peplinear}.

Figure~\ref{belinear} and Figure~\ref{peplinear} illustrate
 visually how one
can, with the help of Eq.~(\ref{linear}), and measurements of the
neutrino absorption and electron scattering rates, 
determine  the 
survival probability $P$ at a given energy.

\section{Neutral Current Experiments}
\label{ncexperiments}

Raghavan, Pakvasa, and Brown\cite{raghavan86} proposed studying the
neutral current 
excitation of individual nuclear levels  
in the same detector in which the $\nu_e$ flux was
measured. As possible targets, they suggested $^{11}$B, $^{40}$Ar, and
$^{35}$Cl, all of which are sensitive to the continuum neutrinos from
$^8$B beta decay in the sun. In
a more recent study, Raghavan, Raghavan, and Kovacs\cite{raghavan93}
proposed using a 4 ton LiF detector with potentially keV energy
resolution to study neutral current and charged current solar neutrino
reactions.  Most recently, Bowles and Gavrin\cite{gaas} have proposed
using neutral current excitations of $^{71}$Ga, $^{69}$Ga, and $^{75}$As
to help diagnose the composition  of the solar neutrino spectrum.  

Let us consider as an especially promising example 
the neutral current excitation of the first
excited state of $^7$Li, which lies $0.478$
MeV above the ground state of $^7$Li. The neutral current excitation
can be represented by the equation

\begin{equation}
\nu ~+~ {\rm ^7Li} ~\rightarrow ~\nu^\prime ~+~ {\rm ^7Li}^*.
 \label{Lincreaction}
 \end{equation}
The reaction can be observed by detecting the $0.478$ MeV
de-excitation $\gamma$-rays.
The energy threshold 
for this reaction is sufficiently low that both the higher-energy
($0.862$ MeV)  
$^7$Be line and  the 
$pep$ neutrinos can excite reaction (\ref{Lincreaction}).  The $pp$
neutrinos and the lower-energy ($0.384$ MeV) $^7$Be neutrinos are not
sufficiently energetic to cause reaction~(\ref{Lincreaction}).
The $^8$B, $^{13}$N, and $^{15}$O neutrinos, as well as $^7$Be and
$pep$ neutrinos,  can all contribute to the
total
observed neutrino excitation of $^7$Li. 

The neutral current matrix element for reaction (\ref{Lincreaction}) 
is large and is 
known accurately\cite{raghavan93} since
the matrix element for reaction (\ref{Lincreaction}) is, by isotopic
spin invariance, the same as the matrix element 
for the observed superallowed decay from the ground
state of $^7$Be to the first excited state of $^7$Li.

If both the neutral current excitation and the charged current
absorption could be  measured for the same neutrino line, then the survival
probability for neutrinos with the energy of the line would be 
 given by the
simple formula 

\begin{equation}
P =  \frac{\sigma_{\rm NC}R_{\rm abs}}
{\sigma_{\rm abs}R_{\rm NC} }. 
\label{Pnccc}
\end{equation}
Here $R_{\rm abs}$ and  $R_{\rm NC}$ are the reaction rates  per target
particle of the charged current (absorption) and neutral current 
processes.
The sensitivity with which the survival probability could be determined
would be 
given by the following relation,

\begin{equation}
\frac{\partial ln P}{\partial ln R_{\rm abs}} = -\frac{\partial ln
P}{\partial ln R_{\rm NC}} = 1.0 .
\label{partialnccc}
\end{equation}

If the neutral current measurement were combined with an
electron-scattering measurement, then the survival probability would be

\begin{equation}
P = \frac{ \sigma_{\rm NC}R_{\rm esc}  -
\sigma_{\rm esc}\left(\nu_{\rm x}\right)R_{\rm NC} }
{ 
\left[\sigma_{\rm esc}\left(\nu_e\right) - 
\sigma_{\rm esc}\left(\nu_{\rm x}\right)\right] R_{\rm NC}
}.
\label{Pncesc}
\end{equation}
The sensitivity with which the survival probability would be 
determined is

\begin{equation}
\frac{\partial ln P}{\partial ln R_{\rm esc}} = -\frac{\partial ln
P}{\partial ln R_{\rm NC}} = \frac{\sigma_{\rm esc}(\nu_e) P + 
\sigma_{\rm esc}(\nu_{\rm x}) (1 - P)}
{\left[ \sigma_{\rm esc}(\nu_{e}) - \sigma_{\rm esc}(\nu_{\rm
x})\right] P } .
\label{partialncesc}
\end{equation}
In the limit in which $P$ is very small, Eq.~(\ref{partialncesc})
shows that the survival probability cannot be  determined by a
combination of a neutral current measurement and an electron scattering
measurement. The physical reason for this indeterminancy, indicated by
the presence of $P$ in the denominator of Eq.~(\ref{partialncesc}), is
that both the neutral current rate and the electron scattering rate
depend only on the neutral current interaction when the survival probability is
very small. 

If all three processes, electron-neutrino scattering, neutrino
absorption, and neutral-current excitation were  measured for the same
neutrino line, then the survival probability would be 
over-determined by
Eq.~(\ref{survival}), Eq.~(\ref{Pnccc}), and Eq.~(\ref{Pncesc}).
The extra constraints could be used as a test of the self-consistency of
the experimental measurements.

Unfortunately, neutral current excitations like that shown in
reaction~(\ref{Lincreaction}) do not register the energy of the neutrino
that causes the interaction. In this respect, neutral current
excitations are similar to radiochemical solar neutrino experiments;
they measure the sum of the reaction rates due to all the neutrino
sources above the energy threshold.  It seems likely\cite{raghavan93}, 
with our current
expectations for the low energy fluxes of solar neutrinos (based upon
the standard solar model and existing neutrino oscillation solutions),
that the neutral current excitation of $^7$Li is dominated by the
higher-energy $^7$Be branch.  However, to verify or improve  these
expectations, additional observational information
must be obtained from neutrino absorption experiments or from
neutrino-electron scattering experiments that can 
 identify the fluxes from individual neutrino sources.

As emphasized by previous authors\cite{raghavan86,raghavan93}, 
the principal
role at present 
of a neutral current excitation experiment is to provide a 
measure of the total neutrino reaction rate, independent of neutrino
flavor, for the entire solar neutrino spectrum.  Since neutral current
excitation experiments cannot be used at present to isolate the
contribution of an individual line, we will not discuss these
excitation experiments further  in
this paper.

\section{Model Independent Tests of Electron Flavor Conservation}
\label{modelindependent}

What can 
can be learned about electron flavor conservation and neutrino parameters
by 
combining the  results, for a neutrino line, of an absorption experiment
and a neutrino-electron scattering
experiment?
In answering this question, we present numerical results for 
$N$,  the normalized ratio of neutrino 
electron scattering rate to neutrino absorption 
rate\footnote{
Equation~(\ref{Ndefinition}) has  exactly the same form as the
expression, Eq.~(\ref{Pnccc}),  for the survival probability as
determined from a neutral current and an absorption measurement.
Everything that we calculate in this section for $N$ could  be
calculated
for the survival probability defined by Eq.~(\ref{Pnccc}).  We
chose to carry out 
our numerical calculations for 
neutrino-electron scattering rather than  neutral
current excitation because neutrino-electron scattering experiments
are currently being developed, whereas there is not yet an advanced
proposal to detect neutral current excitations.},

\begin{equation}
N \equiv 
\frac{\left[\sigma_{\rm abs}\left(\nu_e\right)R_{\rm esc}\right]}
{\left[\sigma_{\rm esc}\left(\nu_e\right)R_{\rm abs}\right]}.
\label{Ndefinition}
\end{equation}
Both $R_{\rm esc}$ and $R_{\rm abs}$ are proportional to the total
neutrino flux created in the sun and therefore the absolute value of
the flux cancels out of the ratio $N$.  
If electron neutrino flavor is conserved, then $N \equiv 1.0$
independent of any solar physics. The quantity $N$ plays much 
the same role
for neutrino-electron  scattering and neutrino absorption as does the
 ratio of neutral current to charged current 
 rates that is a primary goal, for
the higher energy ${\rm ^8B}$ neutrinos,  of the
SNO solar neutrino experiment.

If experimental measurements show that $N$ is different from unity,
then that would be a direct proof that electron flavor is not conserved.
We consider in this section what can be learned from experiemnts with
the $0.862$ MeV $^7$Be and $pep$ lines.

Table~\ref{Ntable} presents the values for $N(^7{\rm Be})$ and
$N({\rm pep})$ that are predicted by the best-fit oscillation
solutions to the four operating solar neutrino experiments.  The
uncertainties indicated represent the 95\% C.L. as defined in \cite{bk96}.
Most of the expected solution space is well separated from the
prediction of electron flavor conservation, although there are
relatively small regions of parameter space, especially for the LMA
and vacuum oscillation solutions,  in which the measured
value of $N$ would be indistinguishable from the value of $1.0$
predicted by electron flavor conservation.

Figure~\ref{Nfigure} shows the allowed region in the 
$N(^7{\rm Be})$ and $N({ pep})$ plane that is consistent with 
the four operating solar
neutrino experiments at 95\% C.L.  
Most of the area that is predicted to be occupied in the $N(^7{\rm
Be})$ and
$N({pep})$ plane is clearly separated from the point in the lower
left hand corner at ($1.0,
1.0$)
that is the standard model prediction.
The upper panel of Figure~\ref{Nfigure} shows the solution space for
the SMA and the LMA MSW solutions and the lower panel shows the
 solution space for the vacuum neutrino
oscillations.

 A priori one might  expect to be able to determine  the two neutrino 
oscillation parameters, $\sin^22\theta$ and $\Delta m^2$, by
measuring the two double ratios $N({\rm ^7Be})$ and $N(pep)$. Unique 
solutions 
are obtainable for the SMA and vacuum oscillation solutions. In these
two scenarios,   the
 ${\rm ^7Be}$ and $pep$ lines  are suppressed differently and the  
relative suppression of the two lines depends strongly on the neutrino
oscillation parameters. However,  for  the LMA solution,
 the two lines are
almost always nearly equally suppressed and there are many pairs of
$\sin^22\theta$ and $\Delta m^2$ for which the suppression of the two
lines is practically the same.
If the LMA MSW solution is assumed to be correct, 
 one cannot in general solve uniquely 
 for the neutrino parameters using just the values of 
$N({\rm ^7Be})$ and $N(pep)$.

How accurately can one determine neutrino parameters by measuring the
 two scattering to absorption ratios?  This question is answered by
Table~\ref{Naccuracy} for MSW SMA oscillations and Table~\ref{7Beratios} for
vacuum oscillations.. The entries in the tables give the range of
solutions for $\Delta m^2$ and $\sin^22\theta$ that are consistent
at 95\% C.L. with the four operating solar neutrino experiments.
If $N({\rm ^7Be})$ and $N(pep)$ are each  measured to an accuracy of $\pm
20$\%, then one can read from Table~\ref{Naccuracy} or
Table~\ref{7Beratios} 
the resulting accuracy with
which $\Delta m^2$ and $\sin^22\theta$ 	and $\Delta m^2$ will be
known. For MSW oscillations (see Table~\ref{Naccuracy}), 
 the characteristic uncertainty  in $\Delta m^2$ would be 
about 10\% and the
characteristic uncertainty in $\sin^22\theta$ would be 
a factor of three or less. For vacuum oscillations (see
Table~\ref{7Beratios}), the mixing angle would be determined well,
typically to an accuracy of order 10\% (although less well in some
regions of parameter space).  The mass difference is not as accurately
determined for vacuum oscillations; the uncertainty indicated by
Table~\ref{7Beratios} can be as large as a factor of two, although
there are some regions of parameter space in which the mass difference
would be very well determined.

\section{Do Sterile Neutrinos Exist?}
\label{sterile}

We discuss in Section \ref{sterileabsplussc}
the modifications in the results previously
presented that are required if sterile neutrinos exist.
On a more theoretical  level, we indicate
 in Section \ref{teststerile} how,
in principle, 
the two solar neutrino lines that arise from $^7$Be electron capture
can be used to test for the existence of sterile 
neutrinos\footnote{We consider a general case in which sterile
neutrinos can couple to $\nu_e, \nu_\mu$, or $\nu_\tau$, and we
consider probabilities $P$ that refer to the net conversion (or
survival) of electron type neutrinos which are created in the sun and
detected on earth.}.  Stimulating previous discussions of sterile
neutrinos in the context of solar neutrino experiments can be found in
references \cite{gaas,bilenky93,Calabresu}.

\subsection{Absorption Plus Electron Scattering Experiments}
\label{sterileabsplussc}

If sterile neutrinos exist, the flux of electron-type neutrinos is
still given by $P\phi$, where the survival probability 
$P = P(\nu_e \rightarrow \nu_e; E_\nu)$ and $\phi$ is the total
flux of neutrinos that are created in the sun.  Thus the rate for the
charged current absorption of neutrinos 
given by Eq.~(\ref{CCrate}) has the same
form whether or not sterile neutrinos exist.  However, the 
total flux of active neutrinos of all types will be reduced  by a
factor $1 - P_{\rm sterile} = 1 - P(\nu_e \rightarrow \nu_{\rm
sterile}; E_\nu)$.  If sterile neutrinos exist,  the
neutrino-electron scattering rate is

\begin{equation}
R_{\rm esc} = \left[\sigma_{\rm esc}(\nu_e) - \sigma_{\rm
esc}(\nu_{\rm x}) \right] P \phi ~+~
\sigma_{\rm esc}(\nu_{\rm x}) \left(1 - P_{\rm sterile} \right) \phi,
\label{sterileescrate}
\end{equation}
which reduces to Eq.~(\ref{escrate}) when $P_{\rm sterile} = 0$.
Combining  Eq.~(\ref{sterileescrate}) with  Eq.~(\ref{CCrate}), the 
survival probability in the presence of sterile neutrinos is
\begin{equation}
P = \frac{\sigma_{\rm esc}\left(\nu_{\rm x}\right)\left(1 - P_{\rm
sterile} \right)R_{\rm abs}}
{\sigma_{\rm abs}R_{\rm esc} -
\left[\sigma_{\rm esc}\left(\nu_e\right) - 
\sigma_{\rm esc}\left(\nu_{\rm x}\right)\right]R_{\rm abs} } .
\label{sterilesurvival}
\end{equation}

Comparing Eq.~(\ref{sterilesurvival}) with  Eq.~(\ref{survival}), we
see that the true survival probability is smaller by a factor of 
$\left(1 - P_{\rm sterile} \right)$ than the survival probability 
inferred by ignoring sterile neutrinos, i.e., 
\begin{equation}
{P = \left(1 - P_{\rm sterile} \right)P_{{\rm no~sterile~\nu's}}}.
\label{sterilescaling}
\end{equation}
This reduction also applies if the survival probability $P$ is
determined by comparing the rates of neutral current excitation and
charged current absorption, as summarized in Eq.~(\ref{Pnccc}).  The
relation given by Eq.~(\ref{sterilescaling}) is physically obvious; it
results from the fact that the survival probability is defined as the
fraction of the total neutrino flux that remains electron-type
neutrinos and that only $\left(1 - P_{\rm sterile} \right)$ of the
total flux is counted by measuring the interaction rates in neutral
current experiments or in neutrino-electron scattering experiments.

The fractional uncertainties  in the inferred survival probability 
can be calculated 
from equations that generalize Eq.~(\ref{partialscabs}), i.e., 

\begin{equation}
- \frac{\partial\, ln P}{\partial\, ln R_{\rm esc}} = + \frac{\partial\, ln
P}{\partial\, ln R_{\rm abs}} = \frac{\left[ \left(
\sigma_{\rm esc}(\nu_e)  -\sigma_{\rm esc}(\nu_{\rm x}) \right)P ~+~
\sigma_{\rm esc}(\nu_{\rm x})\left(1 - P_{\rm sterile} \right) \right]}
{\sigma_{\rm esc}(\nu_{\rm x})\left(1 - P_{\rm sterile} \right)} .
\label{partialsterile}
\end{equation}

\subsection{A Test for the Existence of Sterile Neutrinos}
\label{teststerile}

The relative intensity  of the two neutrino lines produced by electron
capture on $^7$Be is determined by nuclear physics that is independent
of the solar environment.  The ratio of the line strengths, the
so-called branching ratio, has been determined accurately from
laboratory experiments and is \cite{isotopes}

\begin{equation}
\frac{\phi\left(E_2\right)}{\phi \left(E_1\right)} = \hbox{Branching
Ratio} = 0.115\ ,
\label{BR}
\end{equation}
where  
$E_1 = 0.384$ MeV ($10.3$\% of the total flux) and $E_2 = 0.862$ MeV
($89.7$\% of the total flux).
We refer here to the familar laboratory energies of the neutrino
lines; the  energies of the solar lines are
increased by $1.24$ keV and $1.29$ keV, respectively\cite{bahcall94}.

For each of the lines, the total active neutrino flux can be obtained
by measuring the neutrino-electron scattering rate and the
charged-current absorption.  
One obtains from Eq.~(\ref{CCrate}) and Eq.~(\ref{sterilesurvival}) 
\begin{equation}
\left(1 - P_{\rm sterile}\right)\phi = \frac{\sigma_{\rm abs}R_{\rm
esc} - \left[\sigma_{\rm esc} \left(\nu_e\right) - \sigma_{\rm esc}
\left(\nu_{\rm x}\right)\right]R_{\rm abs}}{\sigma_{\rm abs} \sigma_{\rm
esc} \left(\nu_{\rm x}\right)} .
\label{Psterilephi}
\end{equation}

Combining Eq.~(\ref{BR}) and Eq.~(\ref{Psterilephi}), we obtain for
the measured ratio of the total neutrino flux at two different
neutrino energies,

\begin{equation}
\frac{1 - P_{\rm sterile}\left(E_1\right)}{1 - P_{\rm
sterile}\left(E_2\right)} = 0.115
\frac{X\left(E_1\right)}{X\left(E_2\right)}\ ,
\label{OneminusP}
\end{equation}
where
\begin{equation}
X \equiv \frac{\sigma_{\rm abs}R_{\rm esc} - \left[\sigma_{\rm
esc}\left(\nu_e\right) - \sigma_{\rm
esc}\left(\nu_{\rm x}\right)\right]R_{\rm abs}}{\sigma_{\rm abs}\sigma_{\rm
esc}\left(\nu_{\rm x}\right)}.
\label{defX}
\end{equation}
Let
\begin{equation}
\zeta \equiv \frac{\left[\sigma_{\rm esc}\left(\nu_e\right) -
\sigma_{\rm esc}\left(\nu_{\rm x}\right)\right]R_{\rm abs}}{\sigma_{\rm
abs}R_{\rm esc}},
\label{zeta}
\end{equation}
then the fractional uncertainties in the values of $X$ from given
experimental uncertainties can be calculated from
\begin{equation}
\frac{\partial\, ln X}{\partial\, ln R_{\rm abs}} = -\, \zeta \frac{\partial\, ln
X}{\partial\, ln R_{\rm esc}} = \frac{-\,\zeta}{1 - \,\zeta} .
\label{partialX}
\end{equation}

If sterile neutrinos exist, and the probability of their being created
depends upon energy, then the ratio of measured quantities given in 
Eq.~(\ref{OneminusP}) must be different from unity.  If 
$P_{\rm sterile}$ is a constant independent of energy, then the ratio in 
Eq.~(\ref{OneminusP}) will also equal unity.  This latter result
describes the fact that a theory with a constant $P_{\rm sterile}$ 
cannot be distinguished experimentally from a theory
in which all of the solar neutrino fluxes are reduced by a constant
factor. In this very special case of an energy-independent 
$P_{\rm sterile}$,, one would have to 
rely on solar model calculations of the total
neutrino flux in order to determine if sterile neutrinos exist.

In very interesting discussions, Calabresu, Fiorentini, and Lissia
\cite{Calabresu} and Bowles and Gavrin\cite{gaas} 
have pointed out that one can also test for the existence of sterile
neutrinos if one accepts the (robustly calculated)  standard solar
model ratio  of the total flux of $pep$ neutrinos to the total flux of 
$pp$ neutrinos.  In this case, one obtains 
a relation similar to
Eq.~(\ref{OneminusP}) for $pep$ and $pp$  by replacing in
Eq.~(\ref{OneminusP}) the $^7$Be 
branching ratio of $0.115$ with the standard solar model 
branching ratio of $2.4\times10^{-2}$ for $pep$ to $pp$ neutrinos.
  
\section{Summary and Discussion}
\label{summary}

The first three decades of solar neutrino research  concentrated
on continuum energy
spectra. Our goal is to focus additional   
attention on what can be
learned from studying solar neutrino lines.  
 We illustrate what may be observed by assuming the 
correctness of the different neutrino oscillation
solutions that fit the four operating solar neutrino experiments.
For all of our considerations, we assume the existence of experiments
with the excellent energy resolution that is 
necessary to separate solar neutrino
lines from continuum solar neutrino sources and from background events.

We  explore first what neutrino-electron scattering experiments
can tell us about the  MSW and vacuum 
neutrino oscillation solutions to the
solar neutrino problems. 
We find (see Figure~\ref{escallowed}) 
that a measurement of the scattering rate of 
either the ($0.862$ MeV) $^7$Be line 
or the $pep$ neutrino line would, when combined with
the results from the operating experiments, eliminate all but one of
the popular neutrino oscillation scenarios. Which particular solution is
permitted in our simulation 
is, of course, determined by which of the three solutions
(small angle MSW, large angle MSW, or vacuum oscillations) that we
assume is correct.  As is shown in
Figure~\ref{escallowed},  
a measurement of the $pep$ line in addition to the $^7$Be line would
 in many cases provide a significant 
 reduction of the domain of allowed neutrino parameters 
over what is possible by studying only 
the $^7$Be line.

The ``all CNO'' scenario for solar nuclear energy
generation predicts (see Figure~\ref{recoilspectrum}) measurably
higher event rates below $1.5$ MeV
and a markedly 
different shape for the  
electron recoil energy spectrum than would be expected, with or
without neutrino oscillations,  for the standard ``$pp$-dominated'' solar
model description of the  energy generation.
Even without obtaining a high-statistics measurement of the possibly
depleted (by oscillations) $^7$Be neutrino flux, 
a measurement of the electron recoil energy spectrum below $1.5$ MeV
could test  the ``all CNO'' scenario experimentally.

The quantitative predictions of what is expected for
neutrino-electron scattering experiments are summarized in
Table~\ref{snuscenarios}, Table~\ref{bestfit} and Figure~\ref{escallowed}.  These predictions
 can be tested by the BOREXINO\cite{borexino},
HELLAZ\cite{HELLAZ}, and HERON\cite{HERON} 
experiments. 

What can be learned about neutrino properties if both the
charged-current reaction rate (neutrino absorption) and the
neutrino-electron scattering rate are measured?  The short answer is:
 one can
determine the survival probability for electron-type neutrinos at 
the energy of the neutrino line.  Equation~(\ref{survival}) expresses
the survival probability in terms of the measured event rates for the
absorption and the scattering 
 experiments and Eq.~(\ref{partialscabs}) shows how accurately the
survival probability can be determined for specified experimental errors.
If both the scattering and absorption rates are measured, the results
lie along a line in the absorption-scattering plane.  The predicted
range of the solutions for the small and large angle MSW solutions are
well separated from the standard model predictions, as is shown in
Figure~\ref{belinear}a 
for ($0.862$) $^7$Be neutrinos and in Figure~\ref{peplinear}a for $pep$
neutrinos.   Most, but not all, of the vacuum oscillation solutions
that are consistent with the four operating experiments are well
separated from the standard model solution, as shown in Figure~\ref{belinear}b
and Figure~\ref{peplinear}b. For the $pep$ neutrinos, 
the small angle and large angle MSW
solutions are separated from each other in the absorption-scattering
plane.  For the $^7$Be neutrinos, there is some overlap in the predicted
domain for the small and large angle solutions.   

If neutrino oscillations occur, there is a factor of about $5$ 
uncertainty in the expected 
neutrino-electron scattering rates for both the $^7$Be and the $pep$
lines. Table~\ref{bestfit} shows that neutrino oscillation solutions
that are consistent with the four operating experiments permit, at
95\% C.L., the rate for the 862 keV $^7$Be line to be anywhere
 between 22\% and 98\% of the
standard model prediction and the 1.442~MeV $pep$ rate  to lie between
21\% and 98\% of the standard model prediction.  The 384~MeV ${\rm
^7Be}$ line, which may be between 34\% and 100\% of the standard
prediction, is difficult to observe because of the intense background
from the $p$-$p$ solar neutrinos.

Neutral current excitations of individual nuclear levels can, as first
proposed by Raghavan, Pakvasa, and Brown\cite{raghavan86}, provide
important information about the total neutrino flux, independent of
neutrino flavor.  
Like radiochemical experiments, neutral current excitations provide
only one measured number, the total rate due to all neutrino sources.
In order to interpret neutral current
excitations, 
one has to make use of theoretical calculations involving
the standard solar model and the oscillation scenarios.
We describe in Section~\ref{ncexperiments} what can be learned from
neutral current excitation experiments at present and what might be
possible in the future.  We signal out as especially promising for a
future experiment the 
neutrino
excitation of the $0.478$ MeV first excited state of $^7$Li.
The superallowed matrix element for this transition 
is large and is known accurately.
Raghavan, Raghavan, and Kovacs\cite{raghavan93} have suggested that a
practical solar neutrino experiment could be carried out with a 4 ton
LiF detector.  The fluorine
in a LiF detector could make possible a simultaneous 
study at high energy resolution of the higher-energy
$^8$B solar neutrinos\cite{f19}.   

Model-independent tests of neutrino flavor conservation can be carried
out by combining the results of an absorption experiment and a
neutrino-electron scattering experiment for a given neutrino line.
The ratio [see Eq.~(\ref{Ndefinition})] 
of the measured neutrino-electron scattering rate to the measured
neutrino absorption rate, normalized by the interaction cross
sections, must be equal to unity if electron neutrino flavor is
conserved.  Any measured value that is significantly different from
$1.0$ would be a direct proof that electron neutrino flavor is not
conserved. 

Table~\ref{Ntable} presents, for different neutrino oscillation
scenarios,  the best-estimates and the 95\% C.L.
predictions for the normalized ratio of neutrino electron scattering
to neutrino absorption.  For the small mixing angle MSW solution, the
best-estimates are $15.1$ and $11.7$ for the $^7$Be and the $pep$
lines, respectively, an order of magnitude different from what is
predicted by neutrino flavor conservation.  

Figure~\ref{Nfigure} shows for both the $^7$Be and the $pep$ lines 
the wide range of values for the normalized ratio (scattering to
absorption) 
that are consistent
at 95\% C.L. with the results of the four operating solar neutrino
experiments.  Only a  small fraction of the allowed solution space
is close to the region (both normalized ratios equal to unity) that is
implied by electron flavor conservation. Thus a measurement of
neutrino absorption and neutrino electron scattering for one (or both)
of the strong neutrino lines would provide a model-independent
demonstration of electron flavor non-conservation, if  the
neutrino oscillation fits to the results of the 
four operating experiments contain the solution to the solar neutrino
problems.

We show in Section~\ref{sterile}A how the 
results of the previous sections must be modified if there exist sterile neutrinos
 that are coupled to electron type neutrinos. The general result is 
that the true electron neutrino survival probability in the
presence of sterile neutrinos is smaller by a factor of $1 - P_{\rm
sterile}$ than the survival probability inferred by neglecting the
possible existence of sterile neutrinos.  

Do sterile neutrinos exist?
One can in principle carry out a model-independent test for the existence of
sterile neutrinos by combining two experiments for each of the two
$^7$Be neutrino lines.  One knows the branching ratio for the two
lines from laboratory
measurements and this ratio only depends upon nuclear physics.  
Equation~(\ref{OneminusP}) shows that one can detect an
energy-dependent probability for transition to a sterile neutrino by
measuring the absorption and the scattering rate for both of the $^7$Be
neutrino lines. However, it will be difficult to study the lower
energy ${\rm ^7Be}$ line because of the background from $pp$
neutrinos. 
If one accepts as correct the robustly-calculated 
standard solar model ratio of $pep$
to $pp$ neutrino fluxes, then one can apply \cite{Calabresu,gaas} the same 
argument as described here for $^7$Be neutrinos
[and therefore 
Eq.~(\ref{OneminusP})] to test for the existence of sterile neutrinos.

\section*{Acknowledgments}

J.N.B. acknowledges support from NSF grant \#PHY95-13835.  The work of
P.I.K. was partially supported by funds from the Institute for
Advanced Study.  This investigation was initially sparked by a
question asked  by F. Calaprice and by R. Raghavan; the question was:
how much more could BOREXINO   learn about neutrino physics if the
$pep$ neutrinos were measured in addition to the $^7$Be neutrinos?
We are grateful to
R. Eisenstein and E. Lisi for valuable comments and discussions.

%\newpage

\begin{table}

\caption[]{
Experimental results for four operating experiments. The experimental results are
given in SNU for all of the experiments except Kamiokande, for which
the result is expressed as the measured ${\rm ^8B}$ flux in units
of ${\rm cm^{-2}s^{-1}}$ at the earth, assuming the standard model
neutrino shape. The ratios of the 
measured values to the corresponding
predictions in the standard solar model of ref.\protect\cite{BP95} are
also given. The result cited for the Kamiokande experiment assumes
that the shape of the ${\rm ^8B}$ neutrino spectrum is not affected by
physics beyond the standard electroweak model. Here 1~SNU is defined
as $10^{-36}$ interactions per target atom per sec \cite{bahcall69}.\label{snudata} }
\begin{tabular}{l c  c}
Experiment &  Result ($\pm 1 \sigma$) &Reference \\ %\hline
\hline
HOMESTAKE & $2.55 \pm 0.17({\rm stat}) \pm 0.18({\rm syst})$ 
\ SNU&\hfil\cite{chlorine}\hfil\\ %\hline
GALLEX & $77.1 \pm 8.5({\rm stat}) \pm ^{+4.4}_{-5.4} ({\rm syst})$
\ SNU&\hfil\cite{gallex}\hfil\\
%\hline

SAGE & $69 \pm 11({\rm stat}) ^{+5}_{-7}({\rm syst})$ 
\ SNU& \hfil\cite{sage}\hfil \\
%\hline

KAMIOKANDE & [$2.89 \pm ^{0.22}_{0.21}$ ({\rm stat}) $\pm 0.35$ ({\rm syst})] 
$\times 10^6~{\rm cm^{-2}s^{-1}}$ &\hfil\cite{kamiokande}\hfil \\ %\hline
\end{tabular}
\label{expresults}
\end{table}

\begin{table}
\caption[]{Recoil electron event rates in SNU from individual 
neutrino sources 
predicted by different solutions of the solar neutrino problem. The
neutrino oscillation parameters in each solution have been assumed to
be those providing a minimum $\chi^2$ \cite{bk96}. The standard model
 fluxes are from reference ~\cite{BP95}. The threshold energy for recoil
electron was set to zero in the calculations.}

\begin{tabular}{l c c c c c c c}
Solution & $pp$& $pep$&$^7{\rm Be}(0.862)$& $^7{\rm Be}(0.384)$&  $^8{\rm
B}$& $^{13}{\rm N}$& $^{15}{\rm O}$ \\ %hline
SSM      & 6.7E+1&1.5E+0         & 2.7E+1            & 1.0E+0    &
3.9E-1 & 2.8E+0& 3.8E+0  \\
SMA      & 6.5E+1&3.3E-1         & 6.2E+0            & 8.2E-1    &
1.5E-1 & 9.0E-1& 8.6E-1  \\
LMA      & 4.8E+1&8.8E-1         & 1.7E+1            & 7.0E-1    &
1.5E-1 & 1.8E+0& 2.2E+0  \\
VAC      & 4.4E+1&3.6E-1         & 1.9E+1            &3.8E-1
&1.5E-1 & 1.7E+0 & 2.2E+0  \\
CNO      & 2.2E-3&1.2E-4         &1.7E-2             & 9.8E-1
&1.4E-1 & 6.8E+1 & 6.2E+1  \\
\end{tabular}
\label{snuscenarios}
\end{table}

 \begin{table}
\caption[]{Best-fit neutrino oscillation predictions for
neutrino-electron scattering.  The best-fit (and
95\% C.L. limits) are given for the ratio of the rate with neutrino 
oscillations to the rate with the unmodified 
standard solar model flux.   The predicted 
event rates and confidence limits for
neutrino-electron scattering are computed using the techniques of reference \cite{bk96} and the
standard model fluxes of reference \cite{BP95}.}
\begin{tabular}{lcccc}
Scenario&$\chi^2_{\rm min}$&${\rm ^7Be/(^7Be)_{\rm SSM}}$&${\rm
^7Be/(^7Be)}_{\rm SSM}$&$pep/(pep)_{\rm
SSM}$\\
&&(0.862 MeV)&(0.384 MeV)&(1.442~MeV)\\
\hline
SMA&0.3&$0.23^{+0.30}_{-0.01}$&$0.81^{+0.19}_{-0.43}$&$0.22^{+0.11}_{-0.01}$\\
LMA&2.5&$0.62^{+0.14}_{-0.16}$&$0.69^{+0.09}_{-0.10}$&$0.58^{+0.17}_{-0.17}$\\
VAC&2.5&$0.71^{+0.27}_{-0.41}$&$0.39^{+0.55}_{-0.05}$&$0.23^{+0.75}_{-0.02}$\\
\end{tabular}
\label{bestfit}
\end{table}

\begin{table}

\caption[]
{The normalized ratio, $N$, of electron scattering rate to neutrino
absorption rate for the $0.862$ MeV  $^7$Be line 
and for the $pep$ neutrino line. The table entries are the values of $N$ 
that are consistent with the four operating
solar neutrino experiments at the 95\% C.L.  Results are presented
for different neutrino oscillation scenarios. The definition of $N$ is
given in Eq.~(\ref{Ndefinition}).}

\begin{tabular}{l c c c c}
Source & Standard & MSW & MSW & Vacuum  \\
       & Electroweak& SMA & LMA & Oscillations \\
\hline
$^7{\rm Be}$ & 1.0 & $15.1^{+34.6}_{-14.0}$ & $1.21^{+0.30}_{-0.11}$ & 
$1.13^{+1.70}_{-0.12}$ \\
$pep$          & 1.0 & $11.7^{+15.2}_{-10.6}$ & $1.23^{+0.33}_{-0.14}$ & 
$5.78^{+18.7}_{-4.78}$ \\
\end{tabular}
\label{Ntable}
\end{table}

\hglue-1.1in\begin{minipage}{7.5in}
\begin{table}
\squeezetable
\tightenlines
\caption{For the SMA MSW solution, the table gives the 
accuracy with which $N({\rm ^7Be})$ and $N(pep)$ determine 
neutrino prameters. 
The entries give the range of $\sin^2 2\theta $ and $\Delta m^2$ that
are consistent at 95\%  C.L. with the four operating solar neutrino
experiments and for which $N({\rm ^7Be})$ and $N(pep)$ are predicted by the
best-fit SMA solution to be within 20\% of the indicated values.
The top entry
is $\sin^2 2\theta $ (multiplied by $10^3$)
and the lower entry is the difference in the squares of the neutrino
masses (multiplied by $10^6$ ${\rm
eV}^2$).}
\begin{tabular}{@{\extracolsep{-10pt}}cccccccccccc}
${N(pep)} \backslash N({\rm ^7Be})$ & 1.5 & 2.0 & 4.0 & 6.0 & 10.0 & 15.0 & 20.0 & 30.0 & 35.0 & 40.0 & 45.0 \\
\noalign{\hrule\vskip 2pt}
\hskip 11pt 2.0 & $3.5 - 6.3$ & $-$ & $-$ & $-$ & $-$ & $-$ & $-$ & $-$ & $-$ & $-$ & $-$ \\
                & $ 9.8 -  10.3$ & $-$ & $-$ & $-$ & $-$ & $-$ & $-$ & $-$ & $-$ & $-$ & $-$ \\
\noalign{\vskip 3pt}
\hskip 11pt 3.0 & $3.3 - 6.6$  & $3.2 - 7.2$ & $-$ & $-$ & $-$ & $-$ & $-$ & $-$ & $-$ & $-$ & $-$ \\
                & $9.3 - 9.5$  & $8.5 - 9.5$ & $-$ & $-$ & $-$ & $-$ & $-$ & $-$ & $-$ & $-$ & $-$ \\
\noalign{\vskip 3pt}
\hskip 11pt 4.5 & $-$  & $-$ & $3.2 - 8.3$ & $3.3 - 4.4$ & $-$ & $-$ & $-$ & $-$ & $-$ & $-$ & $-$ \\
                & $-$  & $-$ & $7.1 - 7.8$ & $6.8 - 7.1$ & $-$ & $-$ & $-$ & $-$ & $-$ & $-$ & $-$ \\
\noalign{\vskip 3pt}
\hskip 11pt 5.8 & $-$  & $-$ & $3.3 - 8.3$ & $3.3 - 9.1$ & $3.6 - 4.2$ & $-$ & $-$ & $-$ & $-$ & $-$ & $-$ \\
                & $-$  & $-$ & $7.1 - 7.4$ & $6.5 - 7.1$ & $6.0 - 6.3$ & $-$ & $-$ & $-$ & $-$ & $-$ & $-$ \\
\noalign{\vskip 3pt}
\hskip 11pt 7.5 & $-$  & $-$ & $-$ & $4.0 - 9.1$ & $3.6 - 10.0$ & $3.8 - 4.8$ & $4.0 - 4.6$ & $4.4 - 4.8$ & $-$ & $-$ & $-$ \\
                & $-$  & $-$ & $-$ & $6.5 - 6.8$ & $5.9 -  6.3$ & $5.5 - 5.9$ & $5.1 - 5.5$ & $4.8 - 5.0$ & $-$ & $-$ & $-$ \\
\noalign{\vskip 3pt}
\hskip 11pt 9.0 & $-$  & $-$ & $-$ & $-$ & $4.2 - 10.5$ & $4.2 - 10.5$ & $4.2 - 5.5$ & $4.4 - 5.2$ & $4.8 - 5.5$ & $-$ & $-$ \\
                & $-$  & $-$ & $-$ & $-$ & $5.6 -  6.3$ & $5.4 -  5.9$ & $5.1 - 5.5$ & $4.6 - 5.0$ & $4.4 - 4.8$ & $-$ & $-$ \\
\noalign{\vskip 3pt}
\hskip 6pt 11.0 & $-$  & $-$ & $-$ & $-$ & $5.2 - 10.5$ & $4.8 - 11.5$ & $4.8 - 11.5$ & $4.8 - 6.0$ & $4.8 - 6.0$ & $5.0 - 6.0$ & $5.2 - 6.0$ \\
                & $-$  & $-$ & $-$ & $-$ & $5.6 - 5.9$ & $5.1 - 5.8$ & $5.0 -  5.5$ & $4.6 - 5.0$ & $4.4 - 4.8$ & $4.3 - 4.7$ & $4.3 - 4.5$ \\
\noalign{\vskip 3pt}
\hskip 6pt 12.0 & $-$  & $-$ & $-$ & $-$ & $7.2 - 10.5$ & $5.0 - 11.5$ & $5.0 - 11.5$ & $5.0 - 6.6$ & $5.0 - 6.3$ & $5.2 - 6.3$ & $5.2 - 6.3$ \\
                & $-$  & $-$ & $-$ & $-$ & $5.6 - 5.8$ & $5.0 - 5.7$ & $5.0 -  5.5$ & $4.6 - 5.0$ & $4.4 - 4.8$ & $4.2 - 4.7$ & $4.2 - 4.5$ \\
\noalign{\vskip 3pt}
\hskip 6pt 18.0 & $-$  & $-$ & $-$ & $-$ & $-$ & $-$ & $7.6 - 12.0$ & $6.6 - 13.0$ & $6.6 - 13.0$ & $6.6 - 10.5$ & $6.6 - 8.7$ \\
                & $-$  & $-$ & $-$ & $-$ & $-$ & $-$ & $4.6 -  5.0$ & $4.2 -  4.8$ & $4.2 -  4.7$ & $3.9 -  4.6$ & $3.9 - 4.4$ \\
\noalign{\vskip 3pt}
\hskip 6pt 24.0 & $-$  & $-$ & $-$ & $-$ & $-$ & $-$ & $-$ & $8.3 - 12.6$ & $7.9 - 12.6$ & $7.9 - 12.0$ & $7.9 - 12.0$ \\
                & $-$  & $-$ & $-$ & $-$ & $-$ & $-$ & $-$ & $4.0 -  4.4$ & $3.8 -  4.4$ & $3.8 -  4.4$ & $3.9 -  4.3$ \\
\end{tabular}
\label{Naccuracy}
\end{table}
\end{minipage}

\hglue-1.1in\begin{minipage}{7.5in}
\begin{table}
\squeezetable
\tightenlines
\caption{For the vacuum oscillation solution, the table gives the 
acuracy with which $N({\rm ^7Be})$ and $N(pep)$ determine 
neutrino prameters. 
The entries give the range of $\sin^2 2\theta $ and $\Delta m^2$ that
are consistent at 95\%  C.L. with the four operating solar neutrino
experiments and for which $N({\rm ^7Be})$ and $N(pep)$ are predicted by the
best-fit vacuum oscillation
 solution to be within 20\% of the indicated values.
The top entry
is $\sin^2 2\theta $ 
and the lower entry is the difference in the squares of the neutrino
masses (multiplied by $10^{11}$ ${\rm
eV}^2$).}
\begin{tabular}{@{\extracolsep{-10pt}}cccccccc}
${\rm N(pep)} \backslash {\rm N(^7Be)}$ & 1.1 & 1.2 & 1.4 & 1.6 & 2.0 & 2.5 & 3.0 \\
\noalign{\hrule\vskip 2pt}
\hskip 11pt 1.5 & $0.67 - 1.0$ & $0.67 - 1.0$ & $0.67 - 1.0$ & $0.67 - 0.98$ & $0.76 - 0.94$ & $0.85 - 0.94$ & $0.89 - 0.92$ \\
                & $5.4 - 10.4$ & $5.4 - 10.5$ & $5.4 - 10.5$ & $6.1 - 10.5$ & $6.2 - 8.0$ & $6.3 - 7.9$ & $6.3 - 6.5$  \\
\noalign{\vskip 3pt}
\hskip 11pt 2.0 & $0.77 - 1.0$  & $0.77 - 1.0$ & $0.77 - 0.98$ & $0.81 - 0.98$ & $0.88 - 0.97$ & $0.92 - 0.94$ & $-$  \\
                & $5.6 - 10.4$  & $5.6 - 10.5$ & $6.0 - 10.6$ & $6.1 - 10.6$ & $6.2 - 10.6$ & $6.29 - 6.31$ & $-$  \\
\noalign{\vskip 3pt}
\hskip 11pt 2.5 & $0.84 - 1.0$  & $0.84 - 1.0$ & $0.85 - 0.98$ & $0.89 - 0.98$ & $0.94 - 0.97$ & $-$ & $-$ \\
                & $5.7 - 8.4$   & $5.7 - 8.4$   & $6.0 - 8.3$  & $6.1 - 6.2$ & $6.2 - 6.2$ & $-$ & $-$ \\
\noalign{\vskip 3pt}
\hskip 11pt 3.0 & $0.89 - 1.0$  & $0.89 - 1.0$ & $0.89 - 0.99$ & $0.92 - 0.99$ & $-$ & $-$ & $-$ \\
                & $5.7 - 8.4$   & $5.7 - 8.4$  & $6.0 - 8.3$   & $6.1 - 6.2$ & $-$ & $-$ & $-$  \\
\noalign{\vskip 3pt}
\hskip 11pt 4.0 & $0.93 - 1.0$  & $0.93 - 1.0$ & $0.93 - 1.0$ & $0.96 - 1.0$ & $-$ & $-$ & $-$ \\
                & $5.8 - 6.1$  & $5.8 - 6.1$ & $6.0 - 6.2$ & $6.1 - 6.2$ & $-$ & $-$ & $-$ \\
\noalign{\vskip 3pt}
\hskip 11pt 5.0 & $0.95 - 1.0$  & $0.95 - 1.0$ & $0.95 - 1.0$ & $0.98 - 1.0$ & $- $ & $-$ & $-$  \\
                & $5.8 - 6.1$  & $5.8 - 6.1$ & $6.0 - 6.15$ & $6.1 - 6.15$ & $-$ & $-$ & $-$  \\
\noalign{\vskip 3pt}
\hskip 6pt 10.0 & $0.98 - 1.0$  & $0.98 - 1.0$ & $0.98 - 1.0$ & $-$ & $-$ & $-$ & $-$ \\
                & $5.9 - 6.1$  & $5.9 - 6.1$ & $6.0 - 6.1$ & $-$ & $-$ & $-$ & $-$  \\
\noalign{\vskip 3pt}
\hskip 6pt 15.0 & $0.99 - 1.0$  & $0.99 - 1.0$ & $0.99 - 1.0$ & $-$ & $-$ & $-$ & $-$  \\
                & $5.9 - 6.0$  & $5.9 - 6.0$ & $6.0 - 6.05$ & $-$ & $-$ & $-$ & $-$   \\
\noalign{\vskip 3pt}
\hskip 6pt 20.0 & $0.996-1.0$  & $0.996-1.0$ & $1.0 - 1.0$ & $-$ & $-$ & $-$ & $-$ \\
                & $5.9 - 6.0$  & $5.9 - 6.0$ & $6.0 - 6.0$ & $-$ & $-$ & $-$ & $-$ \\
\end{tabular}
\label{7Beratios}
\end{table}
\end{minipage}

\begin{figure} 
\vskip 1cm

\caption[]{
Recoil Electron Energy Spectrum.  The computed 
recoil electron energy spectrum 
is shown for different assumed neutrino production and oscillation
scenarios. 
The vertical arrows  indicate the maximum electron energy
produced by each solar neutrino source.
For the standard solar model with no 
oscillations \cite{BP95}, the 
spectrum is indicated by a solid line.  Assuming the standard model
fluxes are modified by neutrino oscillations, the SMA MSW solution is
indicated by the dotted lines, the LMA MSW solution by the 
line with short dashes, and the VAC oscillation solution is indicated
by long dashes. The dot-dashed line labeled CNO corresponds to the
hypothetical case in which solar energy is derived 
almost completely by CNO
reactions and the 
neutrino fluxes are modified by a SMA MSW
solution \cite{bfk96}.
In actual experiments, the sharp features due to individual lines will
be made somewhat smoother by finite energy resolution.
\label{recoilspectrum} }

\vskip 0.8cm

\caption[]{Allowed Parameter Regions for Four Operating Experiments
plus New Neutrino-Electron Scattering Experiments. The results shown
in the top panel were calculated assuming that the 
best-fit SMA MSW solution is
correct; the middle panel assumes the validity of the LMA solution;
and the lowest panel is based upon the vacuum oscillation solution.
The regions of  $\Delta {\rm m}^2$ and
$\sin^22\theta$ allowed at 95\% C.L. by the four operating
experiments  are shown by solid lines.
Adding a hypothetical measurement of the $0.862$ MeV 
${\rm ^7Be}$ neutrino line equal,
within an assumed $10$\% random error, to the 
value computed using the best-fit neutrino oscillation parameters, 
the dotted curve shows
the allowed regions that would apply for  
the four operating experiments plus the line measurement.
If measurements are made of both the ${\rm ^7Be}$ and the 
${pep}$ neutrino lines, the shaded region applies.\label{escallowed} }

\vskip 0.8cm

\caption[]{
The predicted solution space for $0.862$ MeV 
$^7$Be neutrino-electron scattering
rate versus charged current (absorption) rate.  The indicated
solutions are consistent with the four operating solar neutrino
experiments at the 95\% C.L. The upper panel shows that the SMA and
LMA MSW solutions overlap somewhat in the plane shown, but are well
separated from the predictions of the standard solar model, indicated
by SSM.  The allowed solution space for the vacuum oscillations is
displayed in the lower panel.  These results illustrate the relation
summarized by Eq.~(\ref{linear}).  \label{belinear} }

\vskip 0.8cm

\caption[]{
The predicted solution space for the $pep$ neutrino-electron scattering
rate versus charged current (absorption) rate.  The quantities
displayed are the same as in Fig.~\ref{belinear} except that
Fig.~\ref{peplinear} refers to the $pep$ line. \label{peplinear} }

\vskip 0.8cm

\caption[]{The Allowed Region in the $N(^7{\rm Be})$ 
and $N({pep})$ plane (see text for an explanation of the notation). 
The darkened regions are consistent at the
95\% C.L. with the four operating solar neutrino experiments. The
upper panel shows the allowed solution space for the SMA and LMA MSW
solutions and the lower panel shows the allowed solution space for the
vacuum oscillations.
\label{Nfigure} }

\end{figure}

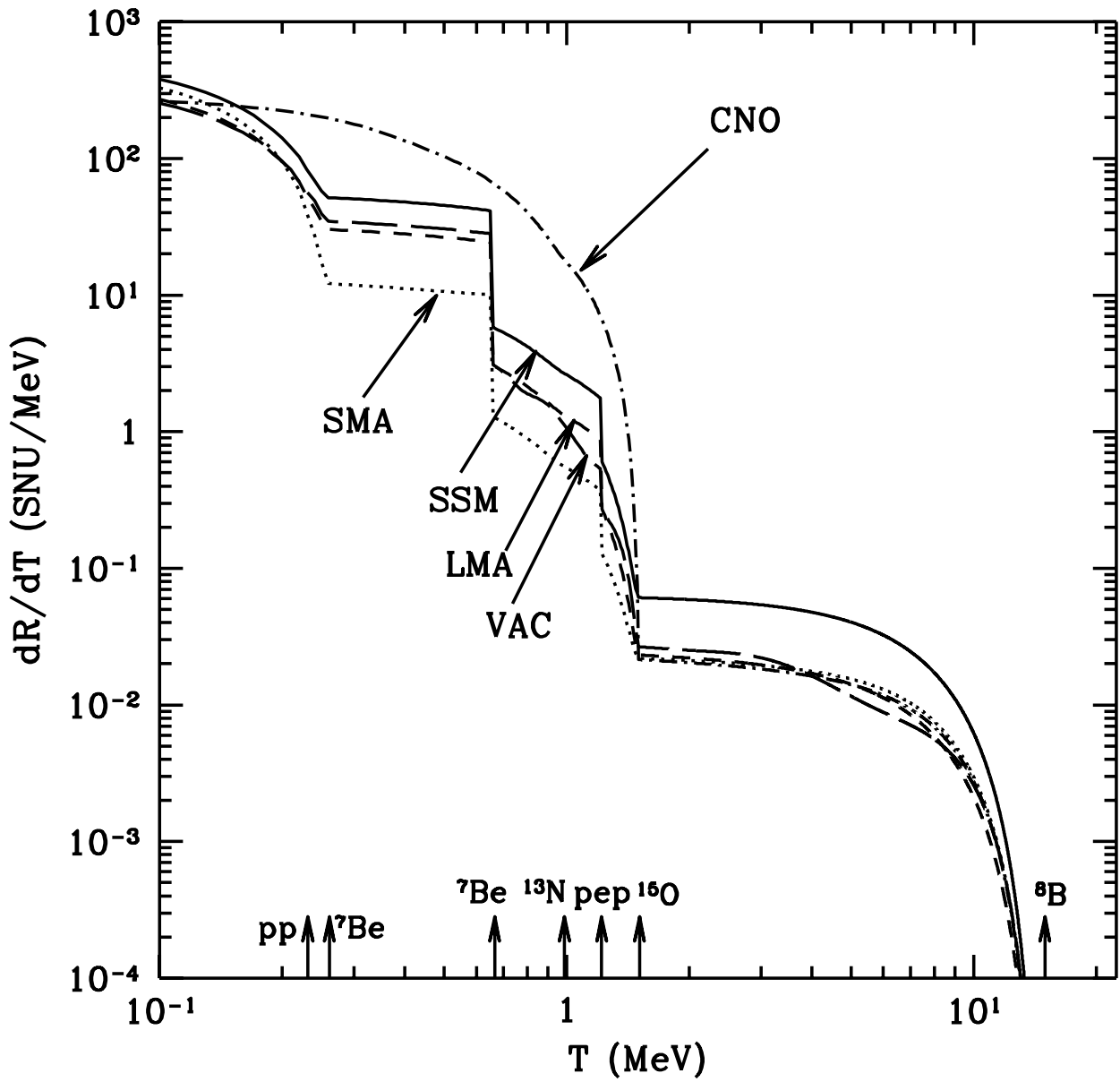
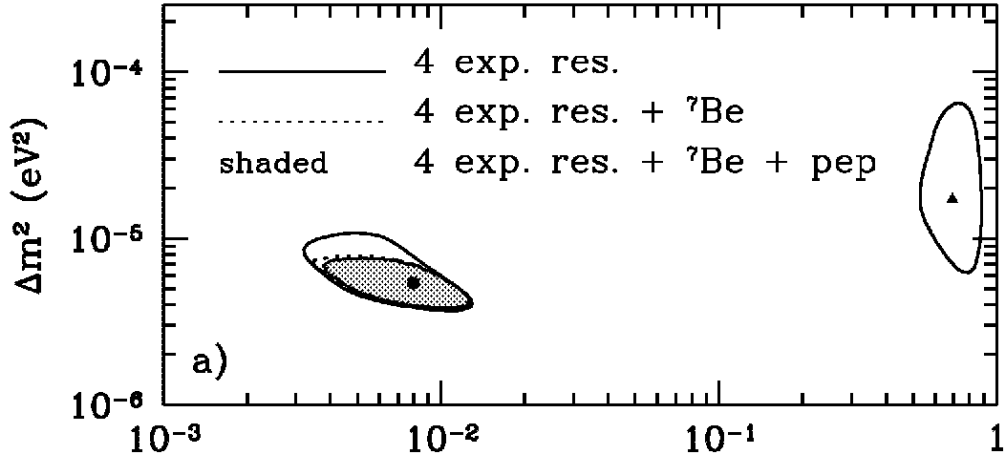
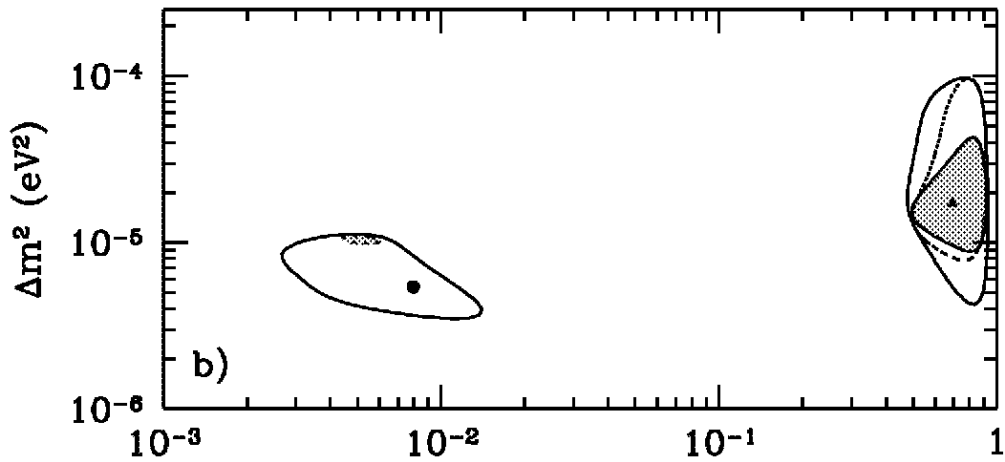
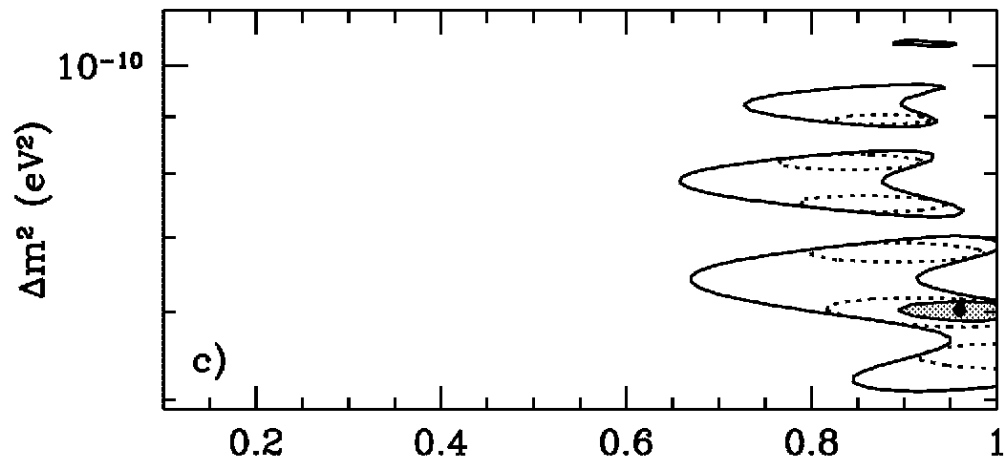
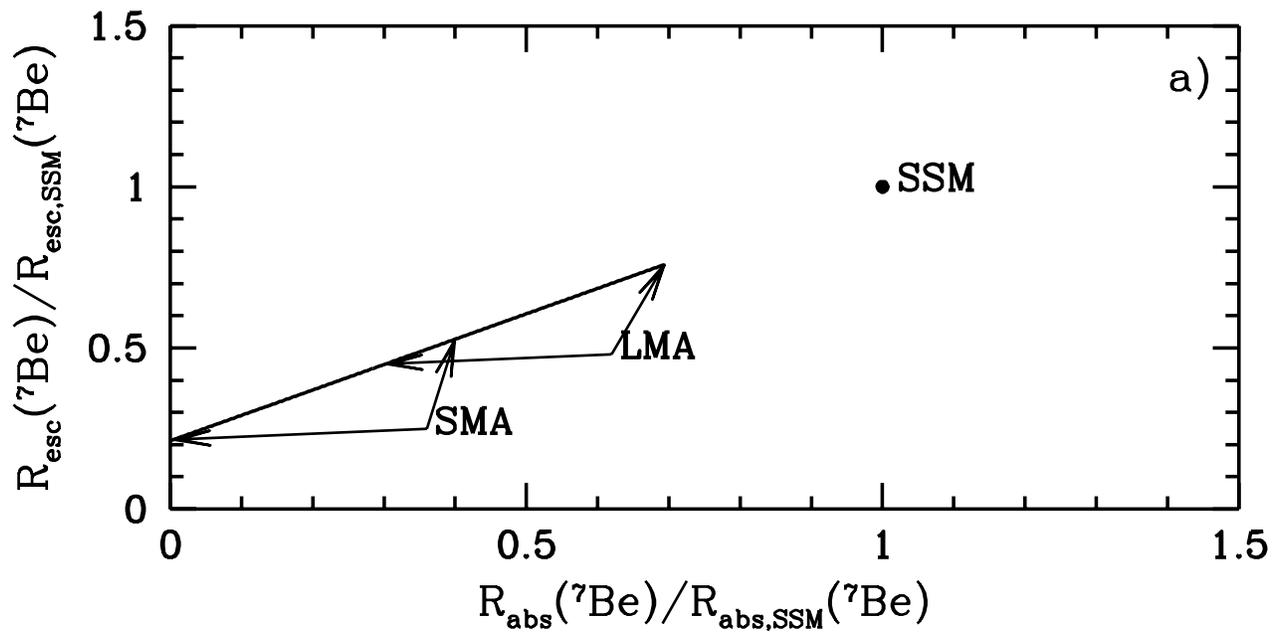
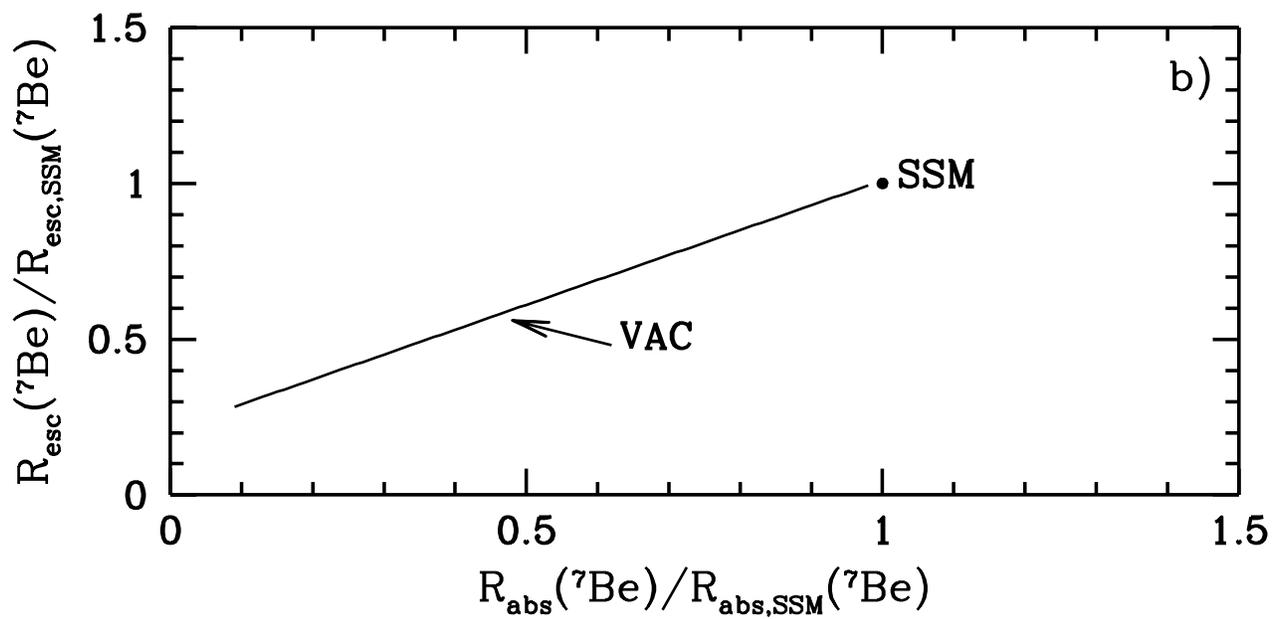
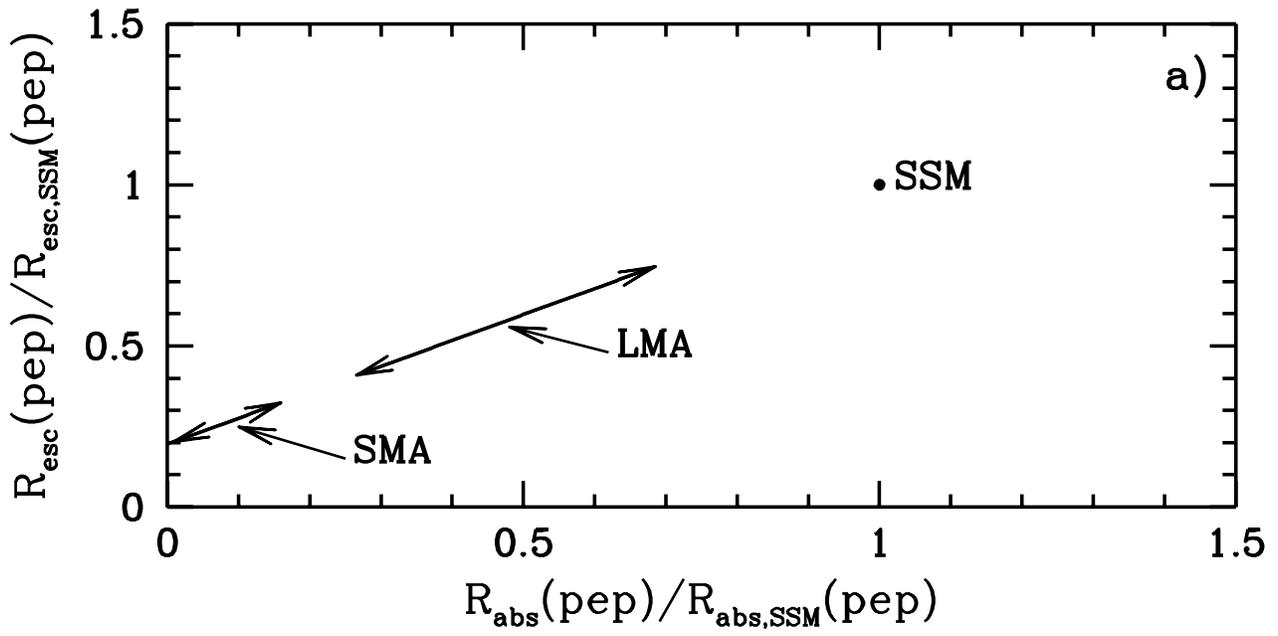
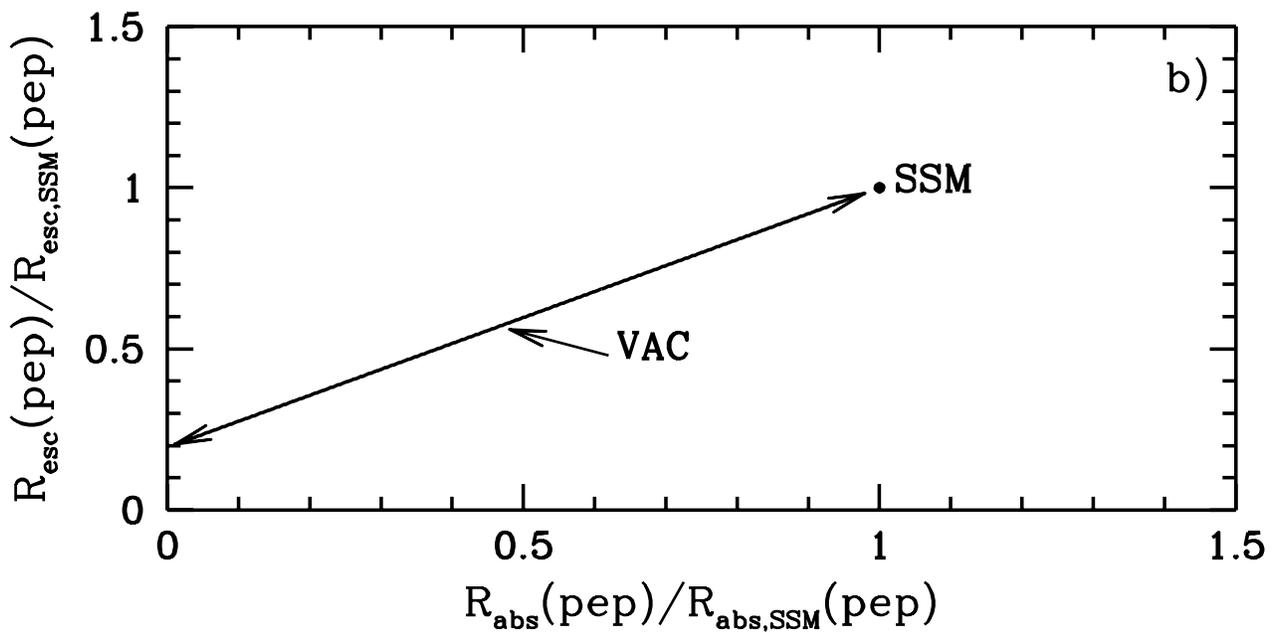
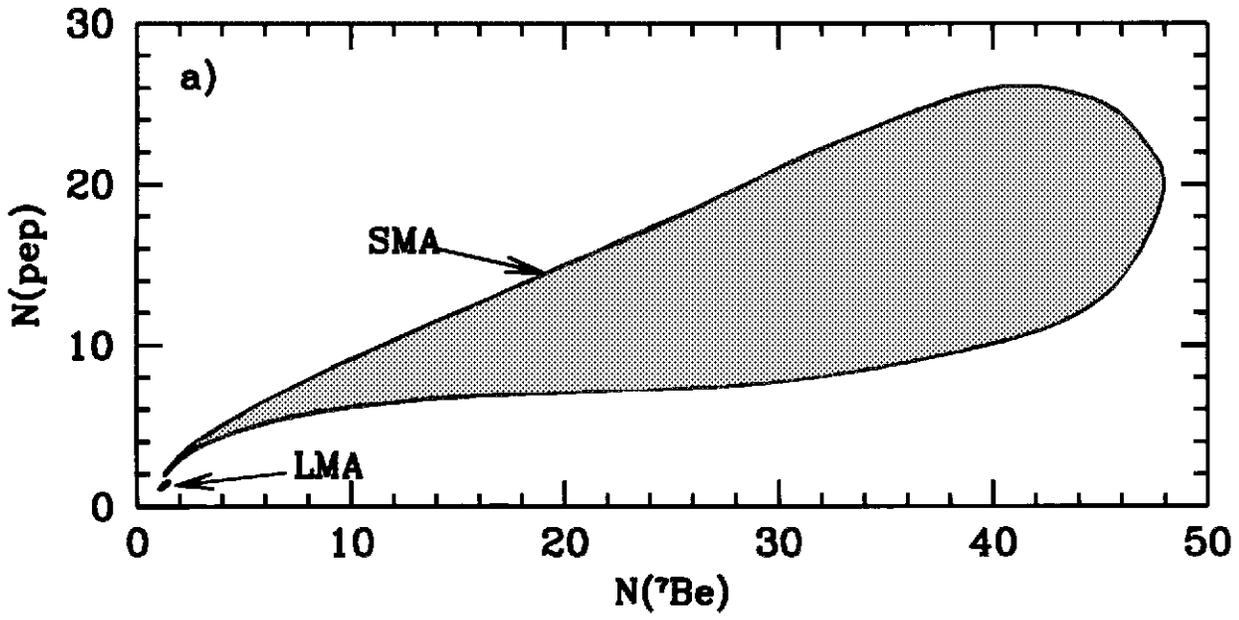
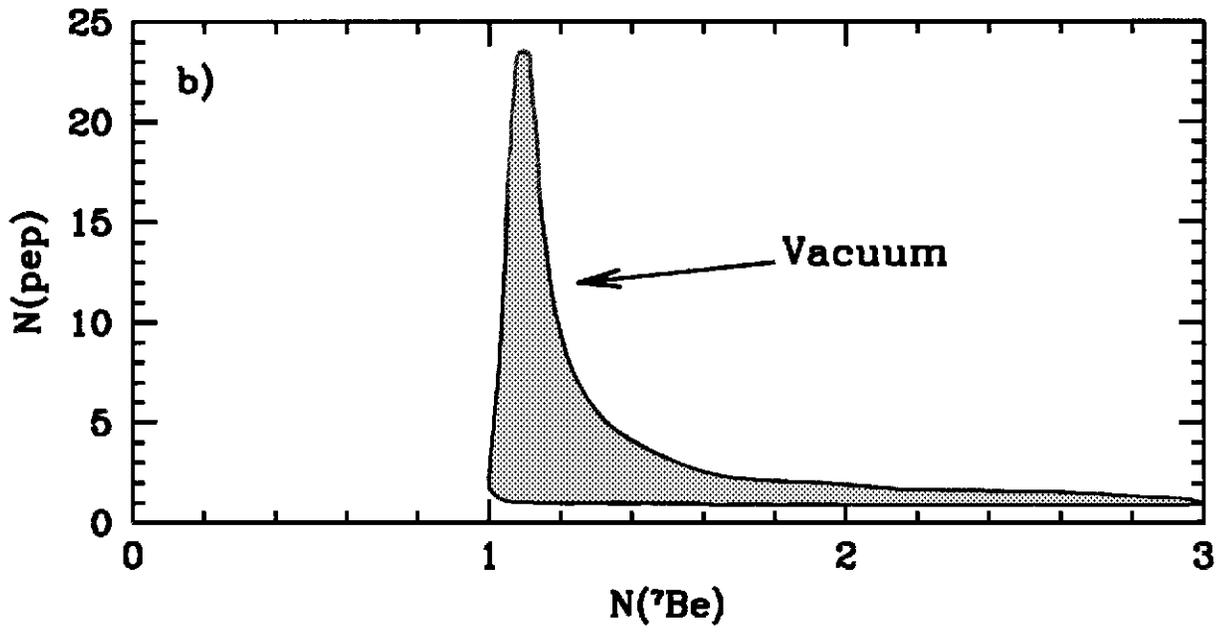
\end{document}